\documentclass[aps,prd,twocolumn,nofootinbib]{revtex4-1}
\usepackage{epsfig}
\usepackage[colorlinks,linkcolor=blue,anchorcolor=blue,citecolor=blue,urlcolor=blue,breaklinks=true]{hyperref}
\usepackage{graphics}
\usepackage{color}
\usepackage{amsmath}
\usepackage[utf8]{inputenc}

\begin{document}
\author{Zhu-Fang Cui$^{1,2,}$}~\email[]{e-mail: phycui@nju.edu.cn}
\author{Shu-Sheng Xu$^{1,2,}$}~\email[]{e-mail: xuss@nju.edu.cn}
\author{Bo-Lin Li$^{3,}$}~\email[]{e-mail: blli@nju.edu.cn}
\author{An Sun$^{3,}$}~\email[]{e-mail: sunan@nju.edu.cn}
\author{Jing-Bo Zhang$^{4,}$}~\email[]{e-mail: jinux@hit.edu.cn}
\author{Hong-Shi Zong$^{1,2,5}$}~\email[]{e-mail: zonghs@nju.edu.cn}

\address{$^{1}$Department of Physics, Nanjing University, Nanjing 210093, China}
\address{$^{2}$State Key Laboratory of Theoretical Physics, Institute of Theoretical Physics, CAS, Beijing, 100190, China}
\address{$^{3}$College of Engineering and Applied Sciences, Nanjing University, Nanjing 210093, China}
\address{$^{4}$Department of Physics, Harbin Institute of Technology, Harbin 150001, China}
\address{$^{5}$Joint Center for Particle, Nuclear Physics and Cosmology, Nanjing 210093, China}

\title{Wigner solution of the quark gap equation}
\begin{abstract}
Solutions and their evolutions of the quark gap equation are studied within the Nambu-Jona--Lasinio model, which is a basic issue for studying the QCD phase structure and locating the possible critical end point. It is shown that in the chiral limit case of the vacuum, chiral symmetry will hold if the coupling strength $G$ is small, then the system only has the Wigner solution at $M=0$. If increasing $G$, two symmetric minima will appear as the positive and ``negative'' Nambu solutions, however, the solution $M=0$ now corresponds to a maximum instead of a minimum of the thermodynamical potential, so is not a physically stable state anymore (we call it ``pseudo-Wigner solution''). Besides, it is shown that as the current quark mass $m$ increases, the pseudo-Wigner solution will become negative, and disappear together with the negative Nambu solution if $m$ is large enough. Similar things happen if we increase the temperature or quark chemical potential $\mu$. Some interesting phenomenon is, from some $\mu$ a second local minimum will show up. As $\mu$ increases gradually, it will be stabler than the Nambu solution, survives even the Nambu solution disappears, and approaches $m$, which are just the features of the Wigner solution we expect.

\end{abstract}


\maketitle

\section{INTRODUCTION}
The fundamental theory of strongly interacting quarks and gluons, Quantum Chromodynamics (QCD), is an important part of the Standard Model of particle physics. However, unlike its cousin Quantum Electrodynamics (QED), QCD has much richer structure, quite special complexity and nonlinear properties in the low energy field, which make two of its most fundamental characters: dynamical chiral symmetry breaking (DCSB) and color confinement, have not been fully understood until now.

In particular, the QCD phase diagram in the temperature ($T$) and quark chemical potential ($\mu$) plane is one of the related open problems that has attracted lots of interests, both theoretically and experimentally. Currently, a popular scenario favors a crossover (or second order for the chiral limit case, where the current quark mass $m$=0) at small $\mu$, and then a first order phase transition for larger $\mu$, with some critical end point (CEP, or tri-critical point for the chiral limit case) links the two regions. Thermal QCD confronts matter at high energy density, as supposed to be prevailing during the early phases of the cosmological expansion, while dense QCD is closely related to neutron star properties. Drastic changes in the early Universe, such as first-order phase transition, can produce a stochastic gravitational wave background~\cite{Aoki:2017aws}, therefore researches of QCD phase transitions are helpful for the investigation of the structure of symmetries in the early Universe (the phase transition associated with the electroweak symmetry breaking is not helpful since not being first order); on the other hand, neutron star mass limit at $2M_\odot$ might support the existence of a CEP~\cite{Alvarez-Castillo:2016wqj}. Alternatively, high energy and high density QCD matter can also be produced in collisions of heavy ions at relativistic energy, over the last two decades, a brief summary of this field could be formulated as the discovery of strongly coupled quark-gluon plasma (sQGP), a near-perfect fluid with surprisingly large entropy-density-to-viscosity ratio~\cite{Shuryak:2014zxa}. Nevertheless, it should be stressed here that QCD inherently has a high energy (and thus a short time) scale, therefore it is not only hard to clarify related questions directly from the first principles of QCD, but also quite difficult to address experimentally.

The search for the position or even the existence of the CEP is also one of the hot topics~\cite{Ayala:2014jla,Eichmann:2015kfa,Inagaki:2015lma,Kovacs:2016juc,Pan:2016ecs,Li:2017zny,Cui:2017ilj}, so does the study about equation of state that provides information on pressure, entropy, energies and other thermodynamic variables of interest~\cite{Bazavov:2017dus,Typel:2017vif,Hanauske:2017oxo}, but thanks to the non-perturbative property of QCD at this region, the results are usually quite model dependant. Essential understanding of these kinds of questions needs us to know different phases and how they change with some parameters (such as $T$ and $\mu$) first, which is usually explored via solutions as well as their evolutions of the quark gap equation. For example, the multi-solution issue is recently discussed with different non-perturbative tools of QCD ~\cite{Zong:2004nm,Chang:2006bm,Williams:2006vva,Wang:2012me,Cui:2013tva,Cui:2013aba,Raya:2013ina,Wang:2016fzr}. Usually, the phase that describing the chiral symmetry breaking system is named as Nambu or Nambu-Goldstone phase, which mainly exists in the low $T$ and low $\mu$ region, and color confinement of quarks and gluons is one of the related key emergent phenomena~\footnote{At present, the relation between chiral phase transition and deconfinement is still obscure, but people strongly think that they are closely related to each other. In this sense, although throughout this work we mainly discuss chiral phase transition, but will not distinguish it with deconfinement very rigorously.}. Oppositely, the phase where chiral symmetry is (partially) restored is often called Wigner or Wigner-Weyl phase, which appears at high $T$ and/or high $\mu$ region, and is then thought to be related to the sQGP state. These two phases corresponds to two distinct solutions of the quark gap equation, usually named as Nambu solution and Wigner solution without doubt. Generally, it is thought that at low $T$ and low $\mu$ region Wigner solution might only exist in the chiral limit case, whether it exists beyond chiral limit is still an interesting and open question.

In this work, we will investigate solutions of the quark gap equation and their properties, especially focus on the Wigner solution. In Section~\ref{cs} we discuss some basic aspects of chiral symmetry and its breaking with the help of the two flavour Nambu-Jona--Lasinio (NJL) model; then in Section~\ref{mtmu}, we discuss the influences of the current quark mass, temperature and quark chemical potential on the solutions of the quark gap equation, especially the behaviours of the Wigner solution; at last, a brief summary is given in Section~\ref{sum}.

\section{chiral symmetry and its dynamical breaking}\label{cs}
Chiral symmetry was known to be an essential feature of the strong interaction before QCD, and it also helped a lot for the development of quark model and QCD itself. Actually, this symmetry is not obvious in the spectrum of hadrons (there are no massless fermions or parity doublets), instead, it only appears in theories with massless fermions, where the fields that describing right- and left-handed particles decouple. QCD Lagrangian possesses an approximate chiral symmetry just because the $u$ and $d$ quarks have much smaller current masses compared to the basic energy scale, $\Lambda_{\rm QCD}$. Correspondingly, the breaking caused by the small but nonzero ``bare'' quark masses (arising from coupling to the Higgs field) is called ``explicit'' chiral symmetry breaking. But, even in the chiral limit case, where in Lagrangian level the chiral symmetry is perfect, the QCD system still does not have chiral symmetry, since the QCD vacuum is not invariant under chiral rotations, and this is usually named as ``dynamical'' (or ``spontaneously'', from different profile) chiral symmetry breaking, which is responsible for about 98\% masses of the luminous Universe! This is an important difference between QCD system and a superconductor or Higgs field: chiral symmetry is global and its currents are not coupled to gauge fields, therefore QCD vacua with different orientations of the order parameter are distinguishable. But now, the intractable question arises: DCSB is mainly related to the non-perturbative properties of QCD, which is just the region that we still cannot understand and solve very well up to now!

In the past, people often regard the QCD vacuum as a condensed state of quark-antiquark pairs (like the condensate of Cooper pairs in a superconductor, or the Higgs vacuum in electroweak theory), the corresponding vacuum expectation value of the quark wave function (often referred to as the quark condensate or chiral condensate), $\langle\bar\psi\psi\rangle$, is then a good indicator of chiral symmetry, since it mixes right- and left-handed quark fields,
\begin{equation}\label{lr}
  \bar\psi\psi=\bar\psi_R\psi_L+\bar\psi_L\psi_R,
\end{equation}
and then is not an eigenvalue of the chirality operator. However, a modern perspective is that the quark condensate can be understood as a property of hadrons~\cite{PhysRevC.82.022201,PhysRevC.85.065202}. No matter what, this means quarks would have interaction with the quark condensate, and then reflect the chiral symmetry and its breaking with their effective masses. Nevertheless, although QCD can be handled in almost the same way as QED when the coupling between its quarks and gluons is small, some non-perturbative tools are still needed in the strong coupling regime. In this work, we will use the NJL model to visualize the chiral symmetry and its breaking via studing the solutions and their properties of the quark gap equation.

It is generally believed that NJL model is a very good laboratory for the investigations of low energy physics of QCD, which works rather well in describing the quark dynamics up to intermediate energies. In NJL model, the Lagrangian is constructed to include the basic symmetries of QCD, while all interaction terms are simplified to be four-body interactions. Usually, its Lagrangian density can be written as (throughout this work we will always work in Euclidean space, under the limit of exact isospin symmetry, namely, $m_{\rm u} = m_{\rm d} \equiv m$, and take the number of flavors $N_{\rm f}=2$, the number of colors $N_{\rm c}=3$),
\begin{eqnarray}
\mathcal{L}&=~&\mathcal{L}_0+G \mathcal{L}_{\rm I}\nonumber\\
&=~&\bar{\psi}\left(i\gamma_{\mu}\partial^{\mu}-m\right)\psi+G\left[\left(\bar{\psi}\psi\right)^2+\left(\bar{\psi}i\gamma_5 \mathbf{\tau}\psi \right)^2\right], \label{eq1}
\end{eqnarray}
where a local, chirally symmetric scalar-pseudoscalar four-point interaction of the quark fields is introduced with an effective coupling strength $G$. Then under the mean field approximation, the effective quark mass $M$ can be determined via the self-consistent gap equation,
\begin{equation}\label{gap}
M=m-2G\langle\bar\psi\psi\rangle.
\end{equation}

We see that $M$ is closely related to $\langle\bar\psi\psi\rangle$, especially in the chiral limit case, where it is then proportional to $\langle\bar\psi\psi\rangle$ with the coefficient $-2G$. In this sense, we can then study the chiral symmetry via solving the gap equation, since $M=0$ means chiral symmetry is unbroken (or restored), while $M\neq0$ means the chiral symmetry is broken. On the other hand, the gap equation is a nonlinear integral equation since $\langle\bar\psi\psi\rangle$ is usually calculated as,
\begin{equation}
\langle\bar\psi\psi\rangle=-\int\frac{{\rm d}^4p}{(2\pi)^4}\mbox{Tr}[S(p)],\label{qq}
\end{equation}
in which $S(p)$ is the dressed quark propagator and the trace is to be taken in color, flavor, and Dirac spaces. Its nonlinearity is an important feature for describing non-perturbative properties of QCD, but may also introduce the possibility for having many solutions, mathematically or even physically.

\begin{figure}
\includegraphics[width=0.45\textwidth]{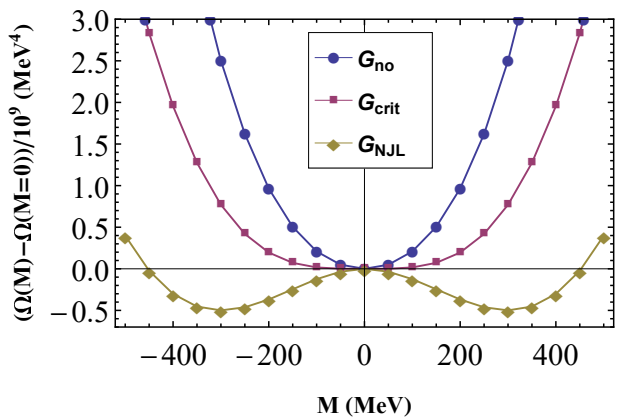}
\caption{(color online) The chiral limit case of $\Omega(M)-\Omega(M=0)$ as a function of $M$ for different coupling strengthes: $G_{\rm no}=3.0\times10^{-6}$ MeV$^{-2}$, $G_{\rm crit}=3.9\times10^{-6}$ MeV$^{-2}$, and $G_{\rm NJL}=5.0\times10^{-6}$ MeV$^{-2}$. Here $\mu=0$, and $T\rightarrow0$.}\label{vacumm}
\end{figure}

To be more clear, let's start from an equivalent way, namely, calculating the thermodynamic potential, by which we can see how many solutions there are, and their positions more directly (actually, it is well known that simple iteration may not give all the possible solutions)
\begin{widetext}
\begin{eqnarray}
\Omega=~G\langle\bar{\psi}\psi\rangle^2- 2N_{\rm c}N_{\rm f}\int^\Lambda\frac{\mathrm{d}^3p}{\left(2\pi\right)^3}{E_p}
-2N_{\rm c}N_{\rm f}\,T\int^\Lambda\frac{\mathrm{d}^3p}{\left(2\pi\right)^3} \left[\ln[1+\mathrm{e}^{-\frac{\left(E_p-\mu\right)}{T}}]+\ln[1+\mathrm{e}^{-\frac{\left(E_p+\mu\right)}{T}}]\right], \label{potential}
\end{eqnarray}
\end{widetext}
where $E_p=\sqrt{{\bf p}^2+M^2}$ is the quark quasi-particle energy, $\Lambda$ is the three-momentum hard cutoff to cure the ultraviolet divergence~\footnote{Here it should be noted that the second integration is actually not divergent, but we still impose the cutoff according to the argument of Ref.~\cite{Cui:2016zqp}.}. Usually, the parameter set $m$=5 MeV, $G=G_{\rm NJL}=5.0\times10^{-6}$ MeV$^{-2}$ and $\Lambda=650$ MeV is used to reproduce pion properties in vacuum. Now let us plot the chiral limit case of $\Omega(M)-\Omega(M=0)$ as a function of $M$ for different coupling strengthes in Fig.~\ref{vacumm}, here we set $\mu=0$ and $T\rightarrow0$.

We can see that, only when the coupling strength is larger than a critical value will DCSB happen, namely, for $G_{\rm no}=3.0\times10^{-6}$ MeV$^{-2}$, there is no DCSB, the vacuum of the system is exactly located at $M=0$, and the gap equation only has this solution; while for $G_{\rm crit}=3.9\times10^{-6}$ MeV$^{-2}$ and larger (for example, $G_{\rm NJL}$), DCSB happens, two symmetric minima appear: the positive one is the normal Nambu solution, while the negative one is also named as (negative) Nambu solution in Ref.~\cite{Wang:2012me}, and the gap equation now has three solution. Here it should be noted that, for the latter case $M=0$ is actually a solution corresponding to a maximum of the potential, so that is only a mathematical solution that looks like the ``Wigner solution'' in the former case, while does not corresponds to a physically stable state at all, so that we will call it ``pseudo-Wigner soluton'' in this work. Also, it can be checked easily that the $M=0$ solution for $G<G_{\rm crit}$ cases does not change qualitatively with $T$ and $\mu$, while will become a positive global minimum beyond the chiral limit, with a value much smaller compared to the normal Nambu solution, and then in some sense can be regarded as the Wigner solution. In other words, if in some model studies that do need two phases in the vacuum, like some bag models, then different coupling strengths should be used. In the following Section~\ref{mtmu}, we will discuss the properties of these solutions for different conditions in detail.

\section{Current quark mass, Temperature and chemical potential effects on the Solutions of the quark gap equation}\label{mtmu}
As we discussed above, chiral symmetry is spontaneously broken by the non-perturbative properties of QCD, how it is affected by parameters, such as current quark mass, finite temperature and quark chemical potential, is then an interesting question of the QCD dynamics.

\begin{figure}
\includegraphics[width=0.45\textwidth]{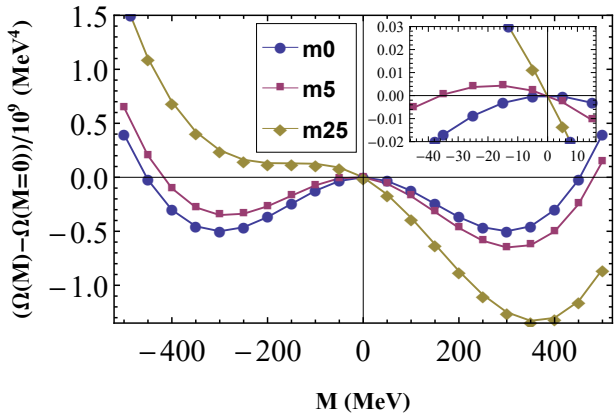}
\caption{(color online) $\Omega(M)-\Omega(M=0)$ as a function of $M$ for different current quark masses: $m0$=0 MeV, $m5$=5 MeV, and $m25$=25 MeV. Here $\mu=0$, and $T\rightarrow0$.}\label{diffm}
\end{figure}

Analysis of a system at non-zero $T$ and $\mu$ can help us to improve the understanding of its vacuum properties. On the one hand, the introduction of some parameters allows us to observe the response of different observables and then can determine some critical parameters; on the other hand, this will provide important input for a quantitative description of the high energy experiments' results, as well as a better understanding of the processes that occurred during the early times of the Universe evolution. First of all, $\Omega(M)-\Omega(M=0)$ as a function of $M$ for different current quark masses is shown in Fig.~\ref{diffm}, here we still set $\mu=0$ and $T\rightarrow0$.

\begin{figure}
\includegraphics[width=0.45\textwidth]{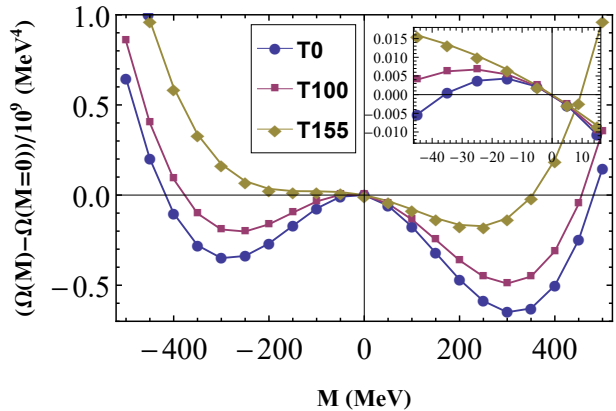}
\caption{(color online) $\Omega(M)-\Omega(M=0)$ as a function of $M$ for different temperatures: $T0\rightarrow0$ MeV, $T100$=100 MeV, and $T155$=155 MeV. Here $m=5$ MeV, and $\mu=0$.}\label{diffT}
\end{figure}

We can see that when beyond the chiral limit, for example $m5=5$ MeV, the pseudo-Wigner solution becomes negative, and the symmetry between the positive and negative Nambu solutions is broken, not only their values, but also their stabilities: the positive Nambu solution becomes the only global minimum, and moves towards its lower right without any qualitative change, while the negative Nambu solution moves to its upper right. Then naturally, there exists a critical value about $m25=25$ MeV, above which the pseudo-Wigner solution and negative Nambu solution will disappear, as shown in the plot. Now to check the temperature effects, we plot $\Omega(M)-\Omega(M=0)$ as a function of $M$ for different temperatures in Fig.~\ref{diffT}, where $m=5$ MeV and $\mu=0$ is used.

We can see from Fig.~\ref{diffT} that, as the temperature increases, the pseudo-Wigner solution moves to its upper left, while the negative Nambu solution moves to its upper right, which is quite similar to their current quark mass effects. However, for the positive Nambu solution quite different trend is observed: it is still always the only global minimum, but now moves towards its upper left, which means the chiral symmetry is restoring step by step. Again, there exist a critical value about $T155=155$ MeV, above which the pseudo-Wigner solution and negative Nambu solution will disappear. By the way, it is not hard to imagine that for the chiral limit case we will have a figure similar to Fig.~\ref{vacumm}, if we regard the $G_{\rm no}$ curve as the high temperature case, $G_{\rm crit}$ curve as the critical temperature where the second order phase transition happens and chiral symmetry is fully restored, while $G_{\rm NJL}$ curve represents lower temperature case.

\begin{figure}
\includegraphics[width=0.45\textwidth]{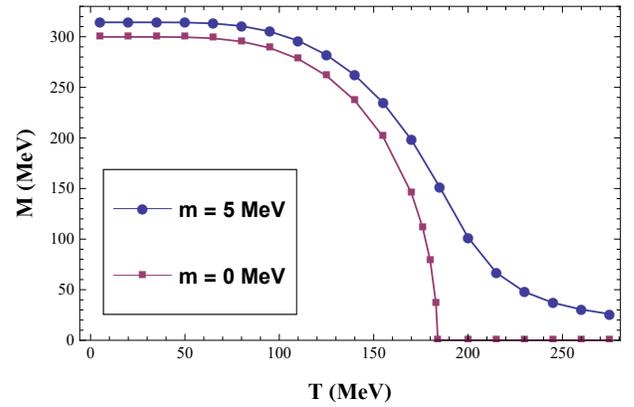}
\caption{(color online)
$M$ as a function of $T$ for different quark masses. Here $\mu=0$.}\label{MT}
\end{figure}

Nevertheless, as shown in Fig.~\ref{MT}, it should be noted that even for very high temperatures, only in the chiral limit case can we have the ``strict'' Wigner solution, while beyond the chiral limit no matter how high the temperature is, the positive Nambu solution will only decreases but never becomes zero, which means the chiral symmetry cannot be fully restored but ``partially'', and then it is hard to define from when can we name the solution as ``positive Wigner'' solution that describes the system where chiral symmetry is partially restored. This phenomenon also exists in other model studies, such as Fig. 1 of Ref.~\cite{Li:2017zny}, which uses Dyson-Schwinger equations. In this sense, QCD susceptibilities, which are the linear responses of the quark condensate to various external fields, can be used as alternative tools~\cite{Cui:2015xta}.

\begin{figure}
\includegraphics[width=0.45\textwidth]{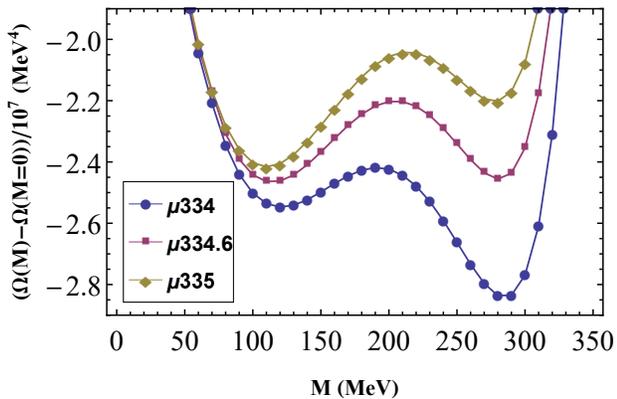}
\caption{(color online) $\Omega(M)-\Omega(M=0)$ as a function of $M$ for different quark chemical potentials: $\mu334=334$ MeV, $\mu334.6$=334.6 MeV, and $\mu335$=335 MeV. Here $m=5$ MeV, and $T\rightarrow0$.}\label{diffmu}
\end{figure}

Further on, we now plot $\Omega(M)-\Omega(M=0)$ as a function of $M$ for different quark chemical potentials in Fig.~\ref{diffmu}, where $m=5$ MeV and $T\rightarrow0$ is used.
Here we only plot a small region, since for one thing, the curve evolution for increasing $\mu$ is qualitatively similar to Fig.~\ref{diffT} (the critical value that the negative Nambu solution and pseudo-Wigner solution disappear is about 303 MeV), and for another, the difference we want to show is small besides change very rapidly, then hard to recognize otherwise.

We can see that, introducing $\mu$ something interesting has happened: a second local minimum may show up! This is not strange since, as we mentioned above, the gap equation is a nonlinear integral equation, it is quite possible that its solution is not unique, at least mathematically. During the calculation, we find that from $\mu\approx333$ MeV, a minimum around $M\approx120$ MeV begins to exist, which is smaller than the positive Nambu solution and then represents a new meta-stable phase, as depicted by the $\mu334$ curve. For $\mu334.6=334.6$ MeV, its potential equals that of the positive Nambu solution, which means coexistence of these two phases and first order phase transition will happen. Then for higher $\mu$ the new phase becomes the stable one and its potential becomes the global minimum, as depicted by the $\mu335$ curve. For $\mu>336$ MeV, the positive Nambu solution will disappear and the new solution will become the only one that survives. As $\mu$ increases, the new solution will decrease and approach $m$=5 MeV gradually, which behaves quite similar to the $T>$180 MeV part of Fig.~\ref{MT}.

In this sense, this is just the Wigner solution we need, which describes a chiral symmetry partially restored state compared to the (positive) Nambu solution, while the pseudo-Wigner solution and negative Nambu solution can only be regarded as mathematical solutions. These results hold until the temperature is high enough and a CEP appears, then there will be no first order phase transition for higher temperatures, instead, the Nambu solution will gradually decrease as in Fig.~\ref{diffT} of the temperature effect, which is then usually named as crossover. The discovery of this real Wigner solution is also useful for locating the CEP precisely. For the chiral limit case, we can get similar conclusions besides the crossover region is replaced by second order phase transition line.

\section{summary and discussion}\label{sum}
Current theoretical studies of the QCD phase structure suggest that a critical end point (CEP) may exist in an experimentally accessibly location, and the search for the position or even the existence of this CEP is then one of the goals of high energy physics, which can also provide important information for the early Universe evolution and compact star properties. But thanks to the non-perturbative property of QCD, this is still a confusing question since the location even existence is quite model dependent up to now.

Fundamental understanding of this question needs us to know how many phases and how they change with some parameters first, which is closely related to chiral symmetry and its dynamical breaking since the strong attraction between quark–antiquark pairs leads to the formation of a scalar quark condensate in the QCD vacuum that spontaneously breaks chiral symmetry, while at high temperature and/or quark chemical potential chiral symmetry may be restored. Usually, this can be investigated by studying solutions of the quark gap equation. In the past, people tend to think that in the chiral limit case of the vacuum, there exist two solutions: the Nambu solution and the Wigner solution, while beyond the chiral limit only the Nambu solution survives. This is surprising and hard to understand.

In this work, we have discussed the solutions and their evolutions with the help of the Nambu-Jona--Lasinio (NJL) model, it is found that in the chiral limit case of the vacuum, if the coupling strength is smaller than a critical value, there will be no dynamical chiral symmetry breaking, and then only the Wigner solution appears at $M=0$, while for larger coupling strengths, two symmetric minima would appear as the positive and ``negative'' Nambu solutions, but now $M=0$ corresponds to a maximum of the thermodynamical potential, so that is rather a mathematical solution which looks like a ``Wigner solution'', but does not corresponds to a physically stable state. Just in this sense, we call it ``pseudo-Wigner solution''.

Furthermore, we have also checked the parameter dependence of the solutions. As the current quark mass increases, the pseudo-Wigner solution will become negative, and the negative Nambu solution moves to its upper right, then above a critical mass the pseudo-Wigner solution and negative Nambu solution will disappear, while the positive Nambu solution becomes the only solution. If we increase the temperature, similar things happen to the pseudo-Wigner solution and the negative Nambu solution, while the positive Nambu solution will decrease gradually, which signals the restoring of the chiral symmetry. If the quark chemical potential $\mu$ is introduced, some interesting things will happen: from some $\mu$ a second local minimum may show up, which represents a new meta-stable phase, for higher $\mu$ it will replace the Nambu solution as the stabler one, and it still exist when $\mu$ is so high that the Nambu solution disappears. Moreover, as $\mu$ increases the new solution will decrease and approach the current quark mass gradually, which is just the characteristic of the Wigner solution we need. Based on these results, it seems that the pseudo-Wigner solution and negative Nambu solution are rather mathematical compared with the positive Nambu solution. On the other hand, if in some model studies that need two phases in the vacuum, like some bag models, based on our model study we conjecture that in principle different coupling strengths should be used.

Last but not least, we need to say that in some sense the NJL model is a ``toy model'' of the Standard Model, so peculiar features of this model need to be confirmed by other QCD-inspired calculations. For one example, readers may question that the positive Nambu solution is also simply a mathematical object, with no physical significance. Yes, in the future people may prove this (just like we propose a new perspective on the Wigner solution in this work), but at present models are still unavoidable, we have to set up some models, fit their parameters with a few quantities, and then try to explain the outputs. For another, readers may also think that QCD's coupling will change in the neighborhood of the critical temperature and/or chemical potential, so it is likely that running couplings will produce results that are quite different from those obtained with a frozen, momentum-independent coupling in this work. We agree but at present, there are little discussion on this subject in literature, the corresponding potential beyond the mean field approximation is also hard to get, and besides, in Ref.~\cite{Cui:2016zqp} we have shown that for some studies, a constant coupling for NJL-like models might be better, as most studies do. Moreover, in Refs.~\cite{Cui:2017ilj,Cui:2016zqp} we also show that different regularization schemes may also lead to quite different results, even qualitatively! In these sense, we see that the need for studies of the QCD phase transition with lattice gauge theory simulations, currently the most powerful approach to QCD, is more urgent than ever.

\acknowledgments
We would like to thank the Referee for many helpful and enlightening comments. This work is supported by the National Natural Science Foundation of China (under Grants Nos. 11805097, 11475085, and 11535005), the Jiangsu Provincial Natural Science Foundation of China (under Grant No. BK20180323), the Fundamental Research Funds for the Central Universities (under Grant No. 020414380051), the China Postdoctoral Science Foundation (under
Grant No. 2015M581765), and the Jiangsu Planned Projects for Postdoctoral Research Funds (under Grant No. 1402006C).

\bibliographystyle{apsrev4-1}
\bibliography{EPJCfinal}

\end{document}